\documentclass[10pt,conference]{IEEEtran}
\IEEEoverridecommandlockouts
\usepackage{cite}
\usepackage{amsmath,amssymb,amsfonts}
\usepackage{algorithmic}
\usepackage{graphicx}
\usepackage{textcomp}
\usepackage{xcolor}
\usepackage{xspace}
\usepackage{listings}
\usepackage{pifont}
\usepackage{makecell}
\usepackage{tcolorbox}
\usepackage{url}
\usepackage{tikz}
\usepackage{balance}

\author{

\IEEEauthorblockN{ Claudio Di Sipio}
\IEEEauthorblockA{\textit{DISIM} \\
	\textit{University of L'Aquila}\\
	L'Aquila, Italy \\
	claudio.disipio@univaq.it}

\and

\IEEEauthorblockN{Juri Di Rocco}
\IEEEauthorblockA{\textit{DISIM} \\
	\textit{University of L'Aquila}\\
	L'Aquila, Italy \\
	juri.dirocco@univaq.it}
\and
\IEEEauthorblockN{ Davide Di Ruscio}
\IEEEauthorblockA{\textit{DISIM} \\
	\textit{University of L'Aquila}\\
	L'Aquila, Italy \\
	davide.diruscio@univaq.it}

\and
\IEEEauthorblockN{Vladyslav Bulhakov}
\IEEEauthorblockA{\textit{DISIM} \\
	\textit{University of L'Aquila}\\
	L'Aquila, Italy \\
	vladyslav.bulhakov@student.univaq.it}
}

\def\BibTeX{{\rm B\kern-.05em{\sc i\kern-.025em b}\kern-.08em
    T\kern-.1667em\lower.7ex\hbox{E}\kern-.125emX}}
\begin{document}

\sloppy

\newcommand*\circled[1]{\tikz[baseline=(char.base)]{\node[shape=circle,draw=black,fill=white,inner sep=0.3pt] (char) {\texttt{\textbf{#1}}};}}

\def\checkmark{\tikz\fill[scale=0.4](0,.35) -- (.25,0) -- (1,.7) -- (.25,.15) -- cycle;} 

%
%

\newcommand*{\ie}{i.e.,\@\xspace}
\newcommand*{\eg}{e.g.,\@\xspace}
\newcommand*{\cf}{cf.\@\xspace}
\newcommand*{\MP}{MAPO\@\xspace}
\newcommand*{\FC}{FOCUS\@\xspace}
\newcommand*{\RM}{README\@\xspace}
\newcommand*{\GH}{GitHub\@\xspace}
\newcommand*{\mA}{\texttt{Llama-2-7b-chat}\@\xspace}
\newcommand*{\mB}{\texttt{Llama-2-13b-chat}\@\xspace}
\newcommand*{\mC}{\texttt{Llama-3-8b-instruct}\@\xspace} 
\newcommand*{\tool}{EvoPlan\@\xspace}
\newcommand*{\CR}{CrossRec\@\xspace}
\newcommand*{\LR}{LibRec\@\xspace}
\newcommand*{\LS}{LibSeek\@\xspace}
\newcommand*{\GR}{GRec\@\xspace}
\newcommand*{\TF}{TopFilter\@\xspace}
\newcommand*{\UM}{UP-Miner\@\xspace}
\newcommand*{\MNB}{MNBN\@\xspace}
\newcommand*{\numPapers}{9,435\@\xspace}
\newcommand*{\numSys}{four\@\xspace}
\newcommand*{\numApps}{2,600\@\xspace}

\newcommand{\library}[1]{\texttt{\hl{\small #1}}}
\newcommand{\librarysmall}[1]{\texttt{\hl{\scriptsize #1}}}

\makeatletter
\newcommand*{\etc}{%
	\@ifnextchar{.}%
	{etc}%
	{etc.\@\xspace}%
}
\makeatother
\newcommand*{\etal}{\emph{et~al.}\@\xspace}
\newcommand\revised[1]{\textcolor{black}{#1}}
\newcommand\old[1]{\textcolor{red}{#1}}

\newcommand{\nb}[2]{
		\fbox{\bfseries\sffamily\scriptsize#1}
		{\sf\small$\blacktriangleright$\textit{#2}$\blacktriangleleft$}
	}

\newcommand\PN[1]{\textcolor{blue}{\nb{Phuong}{#1}}}
\newcommand\JDR[1]{\textcolor{purple}{\nb{Juri}{#1}}}
\newcommand\CDS[1]{\textcolor{red}{\nb{Claudio}{#1}}}
\newcommand\MAX[1]{\textcolor{green}{\nb{MAX}{#1}}}
\newcommand\DDR[1]{\textcolor{orange}{\nb{Davide}{#1}}}

\newcommand{\rqfirst}{\textbf{RQ}: \textit{How effectively can open-source LLMs address popularity bias in TPL recommendations	?}} 

\newcommand{\rqsecond}{\textbf{RQ$_2$}: \textit{To what extent are advanced strategies effective in reducing the popularity bias?}} 




\definecolor{verylightgray}{gray}{0.99}
\definecolor{lightgray}{gray}{0.92}

\lstset{
	basicstyle=\small\ttfamily,
	columns=flexible,
	breaklines=true,
	alsoletter={+},
}

\definecolor{mygreen}{rgb}{0,0.6,0}
\definecolor{mygray}{rgb}{0.95,0.95,0.95}
\definecolor{myred}{rgb}{0.5,0,0}

\lstdefinestyle{JavaStyle} {
	backgroundcolor=\color{verylightgray},   
	commentstyle=\color{mygreen}, 
	breakatwhitespace=false,
	keywordstyle=\color{violet},
	language=Java,
	stringstyle=\color{blue},
	basicstyle=\scriptsize,
	showstringspaces=false
}

\lstdefinestyle{searchstringstyle}{
	basicstyle=\ttfamily\scriptsize,
	captionpos=t,                    
	numbers=none,                    
	numbersep=5pt,                  
	showspaces=false,                
	showstringspaces=false,
	showtabs=false,                  
	tabsize=2,
	frame=single
}

\lstdefinestyle{custom_style}{
    backgroundcolor=\color{white},
    frame=single,            
    rulecolor=\color{black},  
    basicstyle=\ttfamily\footnotesize, 
    breaklines=true,          
    captionpos=b,             
    xleftmargin=0.1cm,        
    xrightmargin=0.1cm,       
}

\lstdefinestyle{xmlstyle}{
  language=XML,
  basicstyle=\footnotesize\ttfamily,
  numbers=left,
  numberstyle=\tiny\color{gray},
  stepnumber=1,
  numbersep=5pt,
  backgroundcolor=\color{white},
  showspaces=false,
  showstringspaces=false,
  showtabs=false,
  frame=single,
  rulecolor=\color{black},
  framexleftmargin=0pt, 
  framexrightmargin=0pt, 
  framesep=0pt, 
  tabsize=2,
  captionpos=b,
  breaklines=true,
  breakatwhitespace=true,
  title=\lstname,
  keywordstyle=\color{black},
  commentstyle=\color{darkgreen},
  stringstyle=\color{black},
  morekeywords={encoding, xmlns, xsi:schemaLocation, xmi:id},
  moredelim=[s][\color{black}]{>}{<},
  moredelim=[s][\color{black}]{\ }{=},
  moredelim=[l][\color{black}]{<},
  moredelim=[l][\color{black}]{</},
  moredelim=[l][\color{black}]{<?},
  moredelim=[l][\color{black}]{/>},
  moredelim=[l][\color{black}]{>},
  escapeinside={(*@}{@*)}
}

\title{Addressing popularity bias in third-party libraries using large language models}

\title{Unveiling the limitation of large language models in third-party libraries recommender systems}

\title{Addressing Popularity Bias in Third-Party Library Recommendations Using LLMs}

\maketitle

\begin{abstract}
    
  	Recommender systems for software engineering (RSSE) play a crucial role in automating development tasks by providing relevant suggestions according to the developer's context. However, they suffer from the so-called popularity bias, \ie the phenomenon of recommending popular items that might be irrelevant to the current task. In particular, the long-tail effect can hamper the system's performance in terms of accuracy, thus leading to false positives in the provided recommendations. Foundation models are the most advanced generative AI-based models that achieve relevant results in several SE tasks. 

This paper aims to investigate the capability of large language models (LLMs) to address the popularity bias in recommender systems of third-party libraries (TPLs). We conduct an ablation study experimenting with state-of-the-art techniques to mitigate the popularity bias, including fine-tuning and popularity penalty mechanisms. Our findings reveal that the considered LLMs cannot address the popularity bias in TPL recommenders, even though fine-tuning and post-processing penalty mechanism contributes to increasing the overall diversity of the provided recommendations. In addition, we discuss the limitations of LLMs in this context and suggest potential improvements to address the popularity bias in TPL recommenders, thus paving the way for additional experiments in this direction. 
\end{abstract}

\begin{IEEEkeywords}
recommender systems, popularity bias, large language models
\end{IEEEkeywords}

\section{Introduction}

While the general concept of fairness, \ie the absence of any bias or prejudice in a given decision-making process, has been widely explored in sensitive domains such as health, crime, or education \cite{fabris_algorithmic_2022,olteanu2019social,chakraborty_software_2019}, \textit{software fairness} is an emerging term that is attracting more and more attention after the rise of AI-intensive models \cite{ferrara_fairness-aware_2023,chakraborty_software_2019,mehrabi_survey_2021} and the related legal concerns underscored by European Union \cite{ai-act}. 

Among the large variety of intelligent systems, recommender systems for software engineering (RSSEs)~\cite{di_rocco_development_2021,robillard_recommendation_2014} are at the forefront in assisting developers in several tasks, spanning from code completion to automated program repair. Among different types of RSSEs, third-party library (TPL) RSSEs provide off-the-shelf software components relevant to the project under development~\cite{9043686,7985674,NGUYEN2019110460,SAIED2018164,LibRec}. 

While previous work demonstrates their accuracy in providing ready-to-use solutions, these systems tend to present frequently seen items~\cite{DBLP:conf/flairs/AbdollahpouriBM19,DBLP:conf/recsys/AbdollahpouriMB19,10.1145/3564284}, thus undermining the novelty of the results. Referred as \emph{popularity bias} in recent literature \cite{deldjoo2022fairness,klimashevskaia_survey_2024,10174041}, this phenomenon is particularly harmful in TPL recommendation, as it can lead to the recommendation of libraries that are not relevant to the current task, thus causing the so-called \emph{long-tail effect} \cite{Anderson:2006:LTW:1197299,4688070} that may hamper the system's performance in terms of accuracy \cite{Vargas_sales_diversity_14}. Although large language models (LLMs) \cite{10.1145/3695988} can be seen as the most advanced intelligent assistant to developers, as showcased in different development tasks \cite{CiniselliCPMAPP22,mastropaolo_studying_2021}, recent research reveal that those models suffer from the long-tail effect in different coding tasks \cite{10298393}. In particular, we aim to answer the following research question:
\begin{quote}
\textbf{RQ:} \textit{How effectively can open-source LLMs address popularity bias in TPL recommendations?}
\end{quote}

To this end, we conduct an ablation study \cite{10.1145/3468264.3468611} by defining six different experimental configurations involving three versions of Llama model~\cite{touvron2023llama} and adopting different strategies to mitigate the popularity bias, \ie few-shots prompt engineering, fine-tuning, and popularity penalty mechanism. 
Our initial findings confirm that the long-tail effect emerges in the recommendations provided by the baseline model, even though fine-tuning and penalty mechanisms can mitigate the popularity bias. Therefore, we see our work as the stepping stone to further investigate this issue and provide more effective solutions to mitigate the popularity bias in RSSEs, \eg employing retrieval augmented generation \cite{10.5555/3495724.3496517} or involving human-in-the-loop solutions \cite{9825861}.

The contributions of our work can be summarized as follows:

\begin{itemize}
\item An initial evaluation on the usage of open-source LLMs to mitigate the popularity bias in TPL recommendations based on six different experimental configurations;
\item A discussion on potential advanced solutions to mitigate the popularity bias in RSSEs;
\item A replication package to foster further research on the topic.\footnote{\url{https://github.com/jdirocco/LLama-2-For-Lib-Rec}}
\end{itemize}

\label{sec:Introduction}

\section{Motivation and Background} 
\subsection{Motivating example}

In AI-based systems, bias can be originated by several factors. Mehrabi \etal proposed a fairness taxonomy~\cite{10.1145/3457607} where three main types of bias have been identified, \ie \textit{Data to algorithm}, \textit{Algorithm to User}, and \textit{User to data}. The first type refers to biases in the training data, \eg unbalanced variables, missing data, or noisy data, while the second type is related to the algorithm itself, \eg the choice of the model, the used hyper-parameters, or the optimization processes. Finally,  \textit{User to Data} bias represents any inherent biases in users that might be reflected in the data they generate. Following the taxonomy, \textit{popularity bias} falls under the \textit{Algorithm to User} category even though we acknowledge that it can be originated by bias in the training data. 



\begin{figure}
    \includegraphics[width=0.9\linewidth]{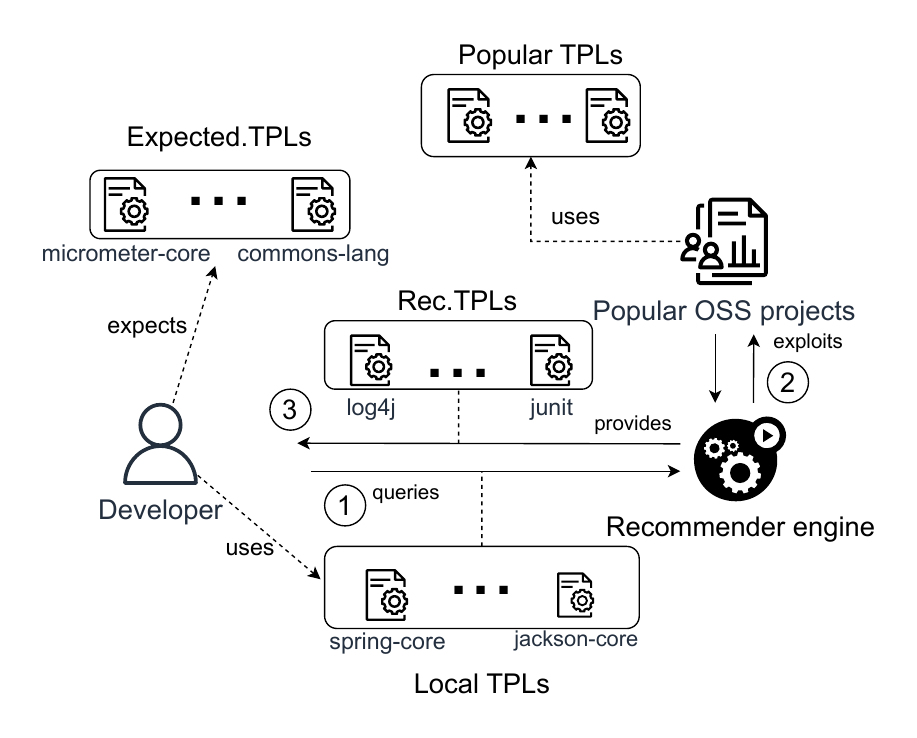}
    \caption{Popularity bias in traditional TPL RSSEs}
    \label{fig:motivation}
\end{figure}

Figure \ref{fig:motivation} represents an explanatory process employing recommender systems for TPLs \cite{NGUYEN2019110460,LibRec}. In the shown example, the user is developing a Web application by relying on \textit{local dependencies}, \ie \texttt{spring-core} and \texttt{jackson-core}. Those dependencies represent the \textit{context} of the TPL recommender, and they may be included in the \textit{query} \circled{1} performed to the system, \ie the \textit{Recommender engine}. During the recommendation phase, the system \textit{exploits} \circled{2} a knowledge base composed of various software artifacts, including OSS projects mined from GitHub. However, those projects use popular TPLs that might not be useful for the current task, thus lowering the TPL recommender's overall accuracy. In the example, the system may \textit{provide} \circled{3} popular libraries, \ie \texttt{log4j} and \texttt{junit} instead of more relevant ones, \ie \texttt{commons-lang3} and \texttt{micrometer-core}, thus reducing the system accuracy. 

Even though existing approaches try to mitigate this issue \cite{9043686}, a recent study reveals that traditional TPLs recommenders still need to address the popularity bias adequately \cite{10174041}. The \emph{Novelty} metric assesses if a system can retrieve libraries in the long tail and expose them to projects~\cite{NGUYEN2019110460}. This increases the possibility of coming across \emph{serendipitous} libraries \cite{Ge:2010_catalog_coverage}, \eg those that are seen by chance but turn out to be useful for the project under development~\cite{di_rocco_development_2021}. For example, there could be a recent library, yet to be widely used, that can better interface with new hardware or achieve faster performance than popular ones. 



On the one hand, project-specific requirements must be considered by the recommender systems to provide more accurate recommendations.
On the other hand, libraries that are well-documented and supported by an active community are more likely to be adopted by developers. This leads to a situation where the most popular libraries are recommended while the less popular ones are overlooked. In summary, recommending only popular TPLs would harm the novelty of the results and
a trade-off between popularity and relevance must be found to provide a more balanced set of recommendations.

\label{sec:Motivation}

\section{Approach}

\begin{figure}
    \includegraphics[width=0.9\linewidth]{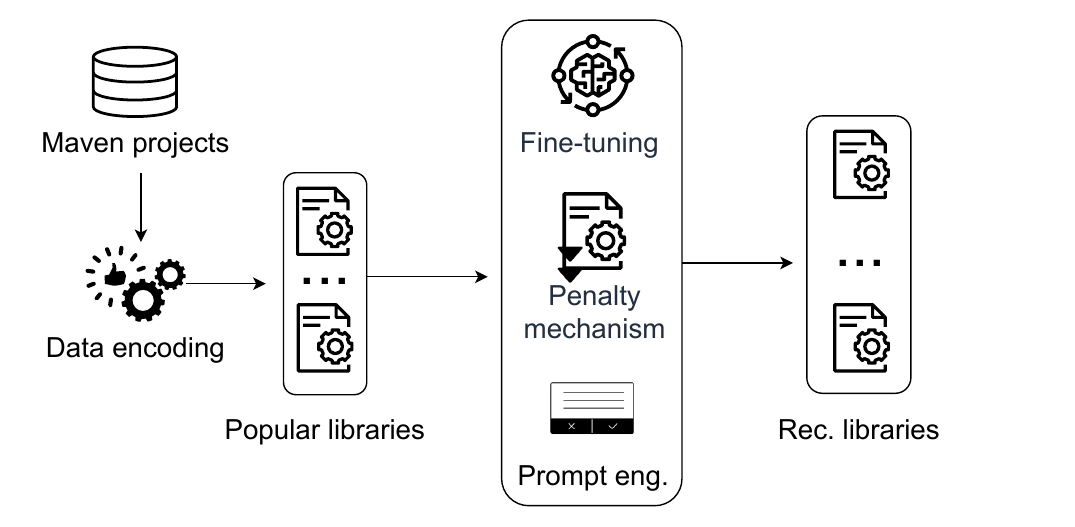}
    \caption{Overview of the proposed approach}
    \label{fig:approach}
\end{figure}

This section outlines the approach employed to investigate the use of open-source LLMs for recommending TPLs, as illustrated in Figure \ref{fig:approach}. The process begins with filtering the original dataset of Java libraries to extract the most popular libraries along with relevant contextual information. Subsequently, the employed foundational model is improved using two complementary strategies: prompt engineering and fine-tuning. To address the long-tail effect in the recommendations, a popularity penalty mechanism is introduced. The following sections provide a detailed explanation of each step.

\subsection{Data encoding}\label{sec:dataEncoding}

To support our analysis, we utilize an existing dataset \cite{NGUYEN2019110460} sourced from \GH projects. We then identify the most popular libraries based on their usage, specifically TPLs that frequently appear as dependencies in other projects. For this purpose, a filtering component is employed to isolate and extract only the popular libraries along with their associated information, such as functionalities, dependencies, README files, and usage scenarios.

For each recommendation session, the system first identifies libraries that are highly popular, specifically targeting the top 20 libraries based on their usage frequency as recorded in the dataset. In addition, each library is annotated with a usage score based on the user interaction data. This allows the system to dynamically adjust recommendations to avoid these highly utilized libraries.

In addition, we collect \RM file for each TPL in the dataset to augment the prompt engineering with contextual information. To avoid token limitation issues during the prompting phase, we summarize the \RM files using an existing approach \cite{doan2023too} that relies on the T5 pre-trained model. In particular, we filter out irrelevant information such as code snippets, URLs, and emojis. We also normalize the text by removing new lines, multiple spaces, and special characters. 

\subsection{Prompt engineering}

After the data curation phase, we conceive a particular prompt that instructs the model to disregard popular libraries and suggest alternative ones. Prompts are dynamically generated based on the current dataset, with parameters adjusted to emphasize features that are unique to less popular libraries, such as dependencies or rare application contexts, as specified in the \RM files. The rationale is to ensure that the model examines a broader array of library options.


In the scope of the paper, we experiment with all two main prompt engineering strategies, \ie zero-shot and few-shot. In addition, we embody the past conversation history in the few-shot learning to enhance the model's ability to generate relevant recommendations.

For each prompt technique, we devise a specific template composed of two main elements, \ie the role and instructions. The former specifies the context and the main objective of the AI assistant. The latter provides a list of fine-grained instructions defined as follows:

\noindent \ding{228}\textbf{IST1:} Ensure recommendations are suitable for the described project context.

\noindent \ding{228}\textbf{IST2:} Focus solely on suggesting libraries that could enhance the project's capabilities or performance.

\noindent \ding{228}\textbf{IST3:} Avoid $popular\_libraries$. 

\noindent \ding{228}\textbf{IST4:} Consider the project's context and existing dependencies to suggest libraries that complement or enhance the current setup.

\noindent \ding{228}\textbf{IST5:} Ignore any code; do not write any code.

\noindent \ding{228}\textbf{IST6:} Provide recommendations exclusively in Maven format as shown in Listing~\ref{lst:output}.

The above mentioned instructions aim to guide the model in recommending lesser-known libraries that are relevant to the project. In addition, we employ the negative prompting technique \cite{kourani_process_2024} to force the LLM to not generate specific content, \eg code snippets. 

The explanatory templates for the zero-shot and few-shots are reported in Listing \ref{lst:zero_shot} and Listing \ref{lst:fewshots}, respectively.

\begin{lstlisting}[style=custom_style, caption={Zero-Shot prompt example.}, label={lst:zero_shot}]
<Role> "As an AI specializing in software library recommendations for Java applications, 
provide recommendations exclusively in Maven format: groupId:artifactId.

<IST1-IST6>


\end{lstlisting}

\begin{lstlisting}[style=custom_style, caption={Few-shots prompt template}, label={lst:fewshots}]
    <role>   
    Example 1:
        Project: <project_name>
        Description: <project_description>
        Existing Dependencies: <list of dependencies>       
    Example 2    ... 
    Example N      
    
   <IST1-IST6>
    \end{lstlisting}

	Meanwhile, Listing \ref{lst:fewshots_hist} shows the template for the few-shot learning with history. In this case, the model is provided with the past conversation history to enhance the recommendation generation process without the list of specific instructions, as we want to evaluate the model's ability to recall past interactions with the user. 

    \begin{lstlisting}[style=custom_style, caption={Few-shots with history.}, label={lst:fewshots_hist}]
	<s>[INST] <<SYS>> <role>
	{{ history }}
	{ if history }
	<s>[INST] Human: {{ input }} [/INST] AI: </s>
	{ else }
	Human: {{ input }} [/INST] AI: </s>
	{ endif }
	
\end{lstlisting}

Listing~\ref{lst:output} shows an example of produced output given in the Maven format.

\begin{lstlisting}[style=custom_style, caption={An explanatory output of the recommendation process.}, label={lst:output}]
	Here is the list in Maven format:
	
	1: org.apache.commons:commons-text
	2: io.jsonwebtoken:jsonwebtoken
	3: com.fasterxml.jackson.module:jackson-module-scalars
	4: org.apache.commons:commons-validator
	5: org.bitbucket.direvtor:javalin
	6: org.jsonwebtoken:jwt-simple
	7: com.github.fge:json-schema-validator
	8: io.requery:jrequery
	9: org.apache.httpcomponents:httpmime
	10: com.nimbusds:oauth2
	
\end{lstlisting}

\subsection{Selected models and Fine-tuning}


Concerning the selected open-source LLMs, we opt for the Llama architecture as its effectiveness in generating content for different tasks has been demonstrated in recent research \cite{radford2019language,brown2020language}. In particular, we experiment with the following models:

\ding{226} \mA\footnote{\url{https://huggingface.co/meta-llama/Llama-2-7b-chat-hf}}: This is part of Meta's Llama 2 family, a collection of pre-trained and fine-tuned generative language models ranging from 7 billion to 70 billion parameters. The 7B chat model is specifically optimized for dialogue-based tasks, fine-tuned using supervised learning and reinforcement learning with human feedback (RLHF). It is designed for assistant-like chat applications and outperforms many open-source models on various benchmarks.

\ding{226} \mB\footnote{\url{https://huggingface.co/meta-llama/Llama-2-13b-chat-hf}}: Similar to the 7B variant, this 13 billion parameters model is fine-tuned for dialogue tasks. It benefits from a larger parameter size, which improves its ability to handle complex natural language tasks with better context understanding and response generation. 

\ding{226} \mC\footnote{\url{https://huggingface.co/meta-llama/Meta-Llama-3-8B-Instruct}}: It is an auto-regressive language model that uses an optimized transformer architecture. The tuned versions use supervised fine-tuning (SFT) and reinforcement learning with human feedback (RLHF) to align with human preferences for helpfulness and safety. 

It is worth mentioning that we opt for small, optimized open-source models to allow easy deployment in local machine. To support the fine-tuning process, we first tokenized the pre-processed datasets as mentioned in Section \ref{sec:dataEncoding}. After the tokenization phase, we adopt the LoRa approach \cite{hu2022lora}, which is a parameter-efficient fine-tuning technique that adjusts specific layers while freezing most of model's original parameters. In particular, we used a rank of 16, a LoRA Alpha of 32, and a dropout rate of 0.05. BitsAndBytes library was used to apply 4-bit quantization, optimizing memory usage using Auto-Tokenizer Python library\footnote{\url{https://huggingface.co/transformers/v3.0.2/model_doc/auto.html}}, with each input truncated or padded to a maximum sequence length of 512 tokens. Table \ref{tab:training_settings} reports the fine-tuning settings in terms of parameters.

\subsection{Popularity penalty mechanism}

As dicussed in Section \ref{sec:Motivation}, existing approaches adopts re-weighting strategies to penalize popular libraries compared to specific ones \cite{gupta_eqbal-rs_2024,9043686}. Similarly, we introduce a popularity penalty mechanism to reduce bias in TPL recommender systems by using data collected from Maven. We define the penalty mechanism as follows:

\begin{equation}
    \text{Penalty Score} = \frac{1}{\text{Popularity Rank} + 1}
\end{equation}

where the popularity rank is the usage of each library collected from Maven. Roughly speaking, we penalize the popular libraries by assigning a lower score to them compared to the less popular ones. The penalty score is then used to adjust the recommendation generation process by reducing the likelihood of recommending popular libraries.

\label{sec:Approach}

\section{Evaluation Materials}

This section discusses the methodology that we used to answer the research question defined in Section \ref{sec:Introduction}. In particular, we aim to evaluate the effectiveness of each defined module in mitigating popularity bias in TPL recommendations.

%
%
%

\begin{figure}
    \centering
    \includegraphics[width=0.4\textwidth]{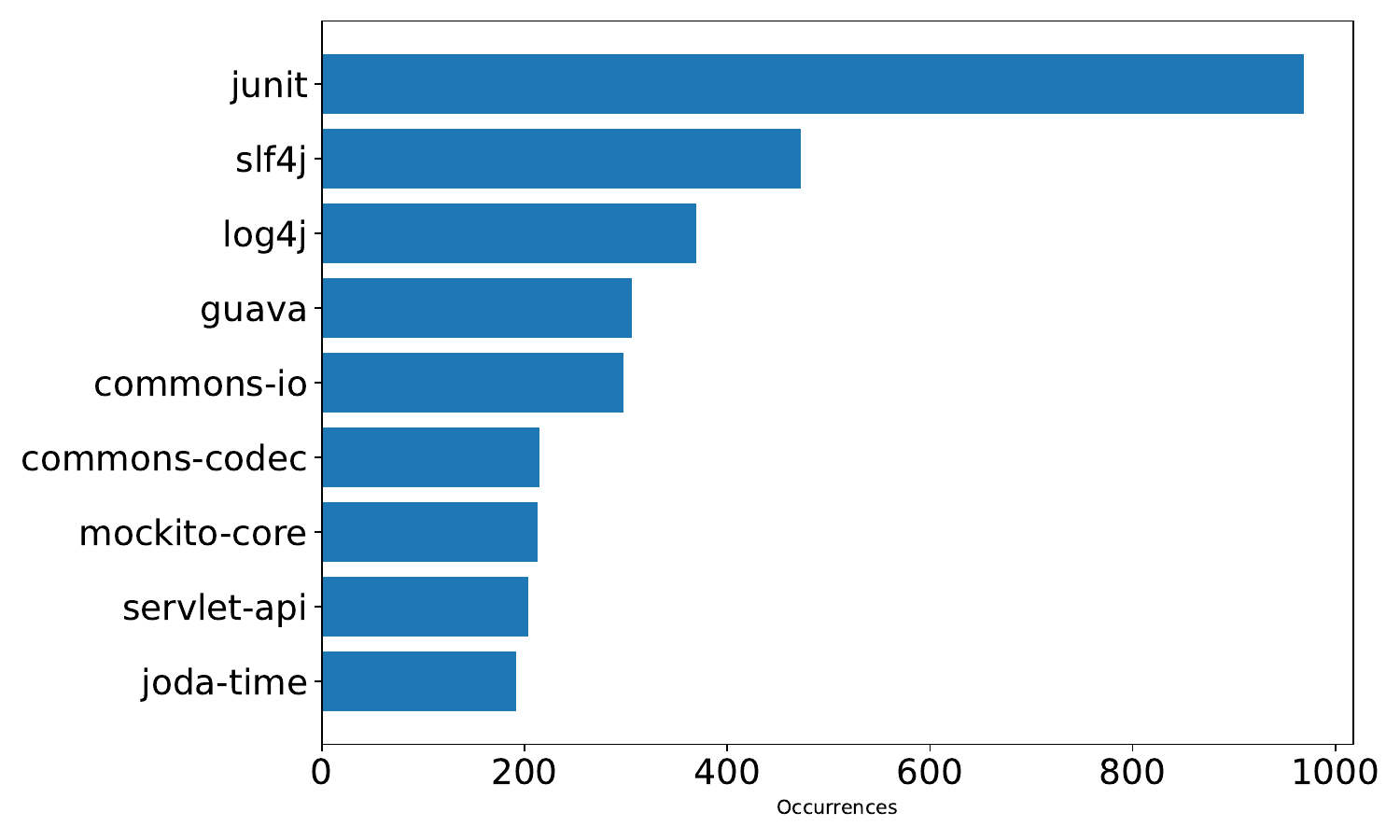}
    \caption{Most popular TPLs in the dataset.}
    \label{fig:popular_libraries}
\end{figure}

\subsection{Metrics}
In the following, we define metrics to assess the system’s performance across relevance diversity and ability to promote less popular libraries.


\smallskip
\noindent
\textbf{Precision and Recall} measure the accuracy of recommendations. Precision (\(P@N\)) is calculated as:

\begin{equation}  
P@N = \frac{\textit{Number of relevant recommended items}}{\textit{Number of recommended items}}
\end{equation}

whereas Recall (\(R@N\)) is defined as:

\begin{equation}
R@N = \frac{\textit{Number of relevant recommended items}}{\textit{Number of items in ground-truth}}
\end{equation}
where the \textit{relevant recommended items} is the set of items in the top-N list that match with those in the ground-truth data.
These metrics evaluate how well system identifies relevant libraries, even when they are not among the most popular.

\smallskip
\noindent
\textbf{F1-Score} balances Precision and Recall, offering a comprehensive view of system’s recommendation accuracy. 
\begin{equation}
F1 = 2 \cdot \frac{P@N \cdot R@N}{P@N + R@N}
\end{equation}

F1-Score is useful in scenarios where maintaining a balance between relevance and reducing popularity bias is crucial.

\smallskip
\noindent
\textbf{Novelty and Diversity} assess how often the system recommends less popular libraries. Novelty measures the ability to introduce new, less-used libraries, addressing the core issue of popularity bias. Diversity evaluates the range of recommendations across different projects, ensuring variety in suggested libraries.

\smallskip
\noindent
\textbf{Catalog Coverage} determines the extent to which the system recommends from the available library catalog. Catalog coverage (\(Coverage@N\)) is calculated as:
\begin{equation}
Coverage@N = \frac{|\bigcup_{p \in P} REC_N(p)|}{|L|}
\end{equation}
where \(REC_N(p)\) represents set of recommended items for project \(p\), and \(L\) is total number of unique libraries available. Higher coverage implies a broader range of recommended libraries.

\smallskip
\noindent
\textbf{Expected Popularity Complement (EPC)} measures the system's ability to recommend libraries that are less popular but still relevant. EPC (\(EPC@N\)) is formally defined as:

\begin{equation}
EPC@N = \frac{\sum_{p \in P} \sum_{r=1}^{N} \frac{rel(p,r) \cdot \frac{1}{1+\log_2(REC_r(p))}}{\log_2(r+1)}}{\sum_{p \in P} \sum_{r=1}^{N} \frac{rel(p,r)}{\log_2(r+1)}}
\end{equation}

where \(rel(p, r)\) is 1 if library at position \(r\) of the top-N list for project \(p\) belongs to ground-truth data, and 0 otherwise. \(REC_r(p)\) reflects popularity of library at position \(r\), ensuring that less popular but relevant libraries are prioritized in recommendations.

\subsection{Ablation study}
\begin{table*}[t]
	\centering
	\caption{Experimental configurations used in the ablation study.}
	\label{tab:configurations}
	\begin{tabular}{|l|l|c|c|c|}
		\hline
		\textbf{Conf.} & \textbf{Model}& \textbf{Prompt technique} & \textbf{Fine-tuning} & \textbf{Penalty Mechanism}  \\ \hline
		C$_1$ & \mA & zero-shot &  \ding{55}& \ding{55}\\ \hline
		C$_2$ & \mB & few-shots &  \ding{55} & \ding{55}\\ \hline
		C$_3$ & \mB & few-shots + history &  \ding{55}  & \ding{55} \\ \hline
		C$_4$ & \mC & few-shots &  \ding{55}& \ding{55}\\ \hline
		C$_5$ & \mC & few-shots &  \ding{55}&  \ding{51}\\ \hline
		C$_6$ & \mC & few-shots &  \ding{51}&  \ding{51}\\ \hline
	\end{tabular}
\end{table*}
This section discusses the ablation study that we conducted with the aim of investigating how
the combination of prompt engineering, fine-tuning, and penalty mechanisms influence the balance between recommendation accuracy and catalog coverage. In particular, we define various configurations of Llama models tested to evaluate their effectiveness in reducing popularity bias within software library recommender systems (see Table \ref{tab:configurations}). These configurations were designed to test different aspects of models’ capabilities and refine our approach based on specific challenges identified in preliminary tests. 

The first three configurations consider the prompts defined in Section \ref{sec:Approach} and two different versions of the Llama 2 model, \ie \mA and \mB. The rationale behind this choice is to investigate how relatively basic LLama models can handle the popularity bias without any specific countermeasures. We consider C$_1$ as the baseline configuration since it considers the basic prompt technique and the \mA model without any enhancement.

Afterward, the two advanced modules, \ie fine-tuning and penalty mechanism, have been tested on \mC in the last two configurations, \ie C$_5$, and C$_6$.  As stated in Section \ref{sec:Approach}, the \mC model exploits RLHF as the underpinning training strategy. Thus, we are interested in understanding how this can affect the popularity bias in TPL recommendations.


For each configuration, we split the dataset into training and testing sets composed of 80\% and 20\% of the data, respectively.
Table \ref{tab:training_settings} summarizes the employed hyperparameters and their corresponding values. 
\begin{table}[h!]
    \centering
    \caption{Fine-tuning settings}
    \label{tab:training_settings}
    \begin{tabular}{|l|l|}
        \hline
        \textbf{HyperParameter} & \textbf{Value} \\
        \hline
        Batch Size & 4 (for both training and validation) \\
        \hline
        Number of Epochs & 3 \\
        \hline
        Learning Rate & 2e-5 \\
        \hline
        Weight Decay & 0.01 \\
        \hline
        Gradient Accumulation Steps & 1 \\
        \hline
        Optimization Algorithm & PagedAdamW (8-bit) optimizer \\
        \hline
    \end{tabular}
\end{table}

\label{sec:Evaluation}

\section{Preliminary Results}
\subsection{\rqfirst}

Table \ref{table:results} shows the results of precision, recall, and catalog coverage for each configuration. Overall, the results demonstrate that the models' performance improved as we introduced more advanced modules compared to the baseline configuration, \ie C$_1$. In particular, using \mA without any enhancement confirms its limitation in handling popularity bias, \ie precision and recall scores are very low. In addition, this configuration achieves only 26\% of catalog coverage, meaning that TPLs are not well diversified. This is due to the limited token size of the model, which prevents it from capturing the full context of the project and generating relevant recommendations. In this respect, we try to enlarge the token size by considering the previous iteration with the model, \ie including the \textit{history} in C$_3$. Even though a slight improvement in all the metrics has been achieved, the overall results remain low, \eg the maximum value of catalog coverage is 36\%.

In contrast, results for the last three configurations show that using the advanced version of the Llama model contributes to improving the overall recommendations even though the combination of different techniques is not yet optimal. On the one hand, the best configuration in terms of accuracy metrics and catalog coverage is C$_3$, where the \mC model is considered with only the few-shots technique. On the other hand, the penalty mechanism achieves the best results in terms of precision, although the catalog coverage decreases to 40\%. Such a negative effect is reduced by introducing the fine-tuning dataset, \ie the catalog coverage and EPC increase up to 55\% and 60\%, respectively. Nonetheless, the recall score is low for all the considered configurations, meaning that even with the conceived advanced techniques, false negatives still have a relevant impact on the recommendations.


\begin{table}[h]
    \centering
    \caption{Results of the ablation study.}    
    \label{table:results}
    \begin{tabular}{|c|c|c|c|c|c|}
    \hline
    \textbf{Configuration} & \textbf{PR@N} & \textbf{REC@N}  & \textbf{F1} & \textbf{Coverage@N} & \textbf{EPC} \\ \hline
    $C_1$ & 0.12 & 0.08 & 0.09 & 26\% & 15\% \\ \hline
    $C_2$ & 0.17 & 0.12 & 0.14 & 30\% & 20\% \\ \hline
    $C_3$ & 0.24 & 0.16 & 0.19 & 34\% & 29\% \\ \hline
    $C_4$ & 0.47 & 0.20 & 0.28 & 58\% & 40\% \\ \hline
    $C_5$ & 0.67 & 0.16 & 0.26 & 40\% & 10\% \\ \hline
    $C_6$ & 0.45 & 0.17 & 0.25 & 55\% & 60\% \\ \hline
    \end{tabular}
    \end{table}

\begin{tcolorbox}[colback=gray!5!white, colframe=black]
\textbf{Answer to RQ:} The ablation study reveals that introducing advanced techniques and prompts has a positive effect on the diversity of the recommendations. However, the recall values are still far from the optimal value, motivating further studies to improve the results. 
\end{tcolorbox}

\subsection{Discussion and improvements}

Despite the advanced techniques proposed in this work, we report that the \textit{popularity bias} is still a relevant issue in open-source LLMs. We acknowledge that further techniques and methodologies are needed to improve the TPLs recommendations. In particular, we plan to investigate the following directions:

\noindent \ding{228} \textbf{Applying bias mitigation algorithms:} Fairness research highlights that a wide range of debiasing algorithms have been successfully applied to traditional ML models \cite{DALOISIO2023103226,10.1007/978-3-030-64881-7_16,ChakrabortyM0M20}. These well-established techniques can be leveraged as a post-processing step to further minimize popularity bias.

\noindent \ding{228} \textbf{Integrating contextual information with RAG:}  Retrieval Augmented Generation (RAG) technique that has been proven to be effective when the contextual information plays a relevant role in SE specific tasks \cite{10.1145/3674805.3695401}. In the context of TPL recommenders, we foresee the usage of this technique to improve the recommendation by selecting unbiased sources of information. For instance, few-shot prompts plus past conversations can be substituted with this advanced technique that can leverage the context directly provided by the user, thus leading to more control over the training data. 

\noindent \ding{228} \textbf{Leveraging user feedback during recommendations:} A relevant field of study in RSSEs is the exploitation of user feedback to improve the recommendations. In the context of our study, we consider the \textit{history} as a source of \textit{implicit feedback}, \ie the model can learn from the previous interactions to improve the recommendations \cite{zhou_braid_2021,ghazimatin_elixir_2021,lomonaco_avalanche_2021}. Future research can investigate how explicit feedback can be introduced in the TPLs RSSE based on LLMs, \eg conceiving different recommendation sessions instead of a single one.
\label{sec:Results}

\section{Threats to validity}

In this section, we discuss the potential threats to the validity of our study and the measures taken to mitigate them.

\medskip
\noindent
\textit{Internal validity} focuses on two primary aspects: the dataset used for the experiments and the prompt engineering techniques. Regarding the dataset, we utilize a state-of-the-art dataset widely adopted in several TPL recommendation systems. Additionally, we identify the most popular libraries in the dataset and leverage their ranking to design the penalty mechanism. As for the prompt techniques, we adhere to well-established guidelines in prompt engineering by employing a consistent prompt across all models and configurations. Furthermore, we enhance the basic prompts by incorporating negative instructions and contextual information from previous iterations with the user.

\medskip
\noindent
\textit{External validity} concerns the generalizability of the study's findings to other contexts, \ie the results may vary if other datasets or LLMs are considered. To mitigate this, we employ a state-of-the-art dataset used by several TPLs recommendation systems, thus ensuring that the obtained recommendations can be compared with those traditional systems. As we focus on open-source LLMs, we opt for Llama 2 and 3, which are widely used in the SE community.

\medskip
\noindent
\textit{Construct validity} refers to the conducted ablation study and its design. While we acknowledge that not all possible combinations have been assessed, we carefully isolate the different components, showing their contributions in terms of well-founded metrics, \ie accuracy, novelty, and diversity. 

\label{sec:Threats}


\section{Related works}
 
\paragraph*{TPL recommendation systems}

\LS~\cite{9043686} is a TPL recommender that provides diversified libraries. The system uses an adaptive weighting mechanism as a post-processing module to neutralize popularity bias by promoting less popular TPLs. The evaluation conducted on a curated Android dataset shows that \LS succeeds in providing a wide range of libraries, thus increasing novelty in the recommendation outcomes. Similarly, Rubei \etal \cite{9825861} investigated the usage of a learning-to-rank mechanism to embody explicit user feedback in TPL recommenders. In particular, the underpinning algorithm is used to re-sort the suggested libraries according to their popularity. They exploit the same dataset used in our study as it contains popular libraries. \LR~\cite{LibRec} works on top of a light collaborative-filtering technique and association mining, retrieving libraries that are used by popular projects.  
Req2Lib \cite{9054865} suggests relevant TPLs starting from the textual description of the requirements to handle the cold-start problem by combining a Sequence-to-Sequence network with a doc2vec pre-trained model.
Similarly, GRec \cite{10.1145/3468264.3468552} encodes mobile apps, TPLs, and their interactions in an app-library graph. Afterward, it uses a graph neural network to distill the relevant features to increase the overall accuracy. Chen \etal \cite{8630054} proposed an unsupervised deep learning approach to embed both usage and description semantics of TPLs to infer mappings at the API level. The model is trained using the information encoded as vectors from 135,127 GitHub projects. 
An approach~\cite{10.1007/s10664-018-9657-y} based on Stack Overflow was proposed to recommend analogical libraries, \ie a library similar to the ones that developers already use. Compared to those approaches, we rely on Llama models to handle the popularity bias in TPL recommendations.


\paragraph*{Analysis of the long tail effect in SE}

In \cite{10298393}, the authors explore several techniques to mitigate the long-tail effect using different pre-trained models in three code-related tasks, \ie API completion, code revision, and vulnerability prediction. The experimental results reveal that the long-tailed distribution has a negative impact on the overall prediction performance, reducing the effectiveness by up to 254.0\% in the examined tasks. Nguyen \etal \cite{10174041} compare three different TPL RSSEs, \ie \LS, \CR, and \LR, to investigate the long-tail effect in TPL recommendations. The authors found that the all the systems are prone to this bias. Lopes and Ossher \cite{10.1145/2858965.2814300} conduct a large-scale analysis on 30,911 Java projects, revealing that they adhere to a long-tail distribution in terms of size, \ie the majority of the projects are small or medium. Borges \etal \cite{7816479} identified four main growing patterns that increase popularity on the GitHub platform. By considering 2,500 top-ranked projects, the conducted study reveals a long-tailed distribution in the examined data. To the our knowledge, our study is the first to investigate the long-tail effect in TPL recommendation systems using LLMs.

\label{sec:RelatedWorks}

\section{Conclusion}
While cutting-edge AI generative models have demonstrated promising results across various software engineering tasks, recent studies highlight a significant challenge: the long-tail effect, which limits diversity in recommendations within software engineering recommendation systems (RSSEs). 

In this paper, we presented an initial investigation into the impact of popularity bias on TPL recommendations generated by open-source LLMs, specifically Llama 2 and Llama 3. To explore this, we applied multiple mitigation strategies aimed at reducing the influence of the most popular libraries in Java projects, including advanced prompt engineering techniques, fine-tuning, and popularity penalty mechanisms. Our results reveal that the long-tail effect is evident in the recommendations produced by the baseline model (Llama 2). However, fine-tuning and penalty mechanisms demonstrate the potential to mitigate this bias in the more advanced Llama 3 model. We hypothesize that incorporating more sophisticated approaches, such as retrieval-augmented generation (RAG) or human-in-the-loop solutions, could further reduce popularity bias.

For future work, we plan to expand our experiments by integrating these advanced solutions, such as bias mitigation techniques, RAG, and user feedback. Additionally, we aim to explore open-source models optimized for code-related tasks, such as CodeMistral and CodeLlama. Finally, we will broaden the dataset to include a larger number of Java projects and extend the analysis to other programming languages that heavily rely on TPLs, such as Python and JavaScript.
\label{sec:Conclusion}

\section*{Acknowledgments}
This work was partially supported by the following Italian research projects: EMELIOT (PRIN 2020, grant n. 2020W3A5FY) and TRex SE (PRIN 2022, grant n. 2022LKJWHC), MATTERS project, funded under the cascade scheme of the SERICS program (CUP J33C22002810001), Spoke 8, within the Italian PNRR Mission 4, Component 2, and the FRINGE project (PRIN 2022 PNRR, grant n. P2022553SL).

\balance
\bibliographystyle{IEEEtran}
\bibliography{main}

\end{document}